# The Professional Challenges of Industrial Designer in Industry 4.0

Meng LI
Industrial Design Department
Xian Jiaotong University
Xi'an, China
limeng.81@mail.xjtu.edu.cn

Yu ZHANG
Industrial Design Department
Xian Jiaotong University
Xi'an, China
zy_xjtu@163.com

Leshan LI
Industrial Design Department
Xian Jiaotong University
Xi'an, China
lileshan2@sina.com

*Abstract*—The Industry 4.0 refers to a industrial ecology which will merge the information system, physical system and service system into an integrate platform . Since now the industrial designers either conceive the physical part of products, or design the User Interfaces of computer systems, the new industrial ecology will give them a chance to redefine their roles in R&D work-flow. In this paper we discussed the required qualities of industrial designer in the new era, according to an investigation among Chinese enterprises. Additionally, how to promote these qualities though educational program.

*Keywords-industry 4.0; industrial designer quality; educational program of design; PHA education model; humanity quality*

## I. THE CHALLENGS FROM NEW INDUSTRIAL REVOLUTION

Since March 25th, 2015, Premier Li Keqiang proposed to promote the implementation of "China Manufacture 2025", which is the ten-year development guideline made by the Ministry of Industry and Information Technology (MIIT for short) for the future Chinese industry, "Internet +" is becoming a hot keyword for the researchers' and entrepreneurs' circle. The "Internet +" represents a trend of Chinese industry reform, in order to handle the "double pressure" from highly industrialized countries like the Germany and the US, as well as the developing industry areas, like the Vietnam and the Africa [1]. Many low-end OEM and ODM segments in China had already transferred to low-labor-cost area. But many of the high-end segments of industry are still dominated by Germany and US, and maybe enforced by their industrial strategies like "Industry 4.0" and "Re-industrialization" . During the new round of Chinese industry reform, the main goal of the manufacture companies is to redefine the their positions in the value chain, which needs benefiting from the fast growing information technologies and industry in China[2]. On May 19, the State Council released a ten-year blueprint "China Manufacture 2025" for conducting Chinese manufacturing development, in which the primary indexes are innovation ability, qualitative efficiency, Information and Computer Technologies (ICT for short) application and industrialization integration, and green growth[3].

### A. A New Industrial Revolution

The concept of "forth industrial revolution" is derive from the German national high-tech strategy that released in 2012 [4]. The "Industry 4.0" is a roadmap which suggested a future perspective of intelligent industrial development in Germany. The aim of this strategy are keep the German industry's world leadership in order to cater impacts from the fast industrial development of emerging economies, dynamic market, growing diversity of user needs and new emerging technologies. The industry 4.0 bases on the mature automation technique and developed embedded system in Germany [5]. Meanwhile, the GE introduced "industrial internet" project in the US, as the third wave after internet revolution, so that the productivity will grow 1-1.5% annually to promote the income level 25-40% in next two decades[6]. The trend of merging the information techniques with manufacture system is not using IT as a tool supporting the production, but as an intelligent machine grid to rebuild technique chain, value chain and innovation chain. Unlike industrialized countries such as Germany and US, the big challenge for Chinese industry is to reform way of industrial development, so as to reduce the consumption of natural resource, environmental contamination and over production, while improve the quality, efficiency and sustainability of development. The ICT sector is the most active part of Chinese industry, so "Internet+", "Intelligent Manufacture" and "China manufacture 2025" are all expect to renovate Chinese manufacture system with this fresh driver[1]. The developed and developing areas both view it a good chance to promote national manufacturing competency. As Manufacturing is the key driver of innovation, R & D and productivity growth, the EU is planning to enforce the innovation in manufacturing though a framework program - Horizon 2020 to counter the de-industrialization trend after mortgage crisis [7]. The low-growth countries like Japan expects to regain their leadership in manufacturing industry by developing systematized design and manufacturing technology to augment human expertise [8]. The focus of manufacturing innovation is not on fast productivity growth, but on the stable growth of operational efficiency of manufacturing system as well as the development of new business model, services and products. The components of this new industrial revolution are: Cyber-Physical System (CPS for short), Internet of Things, Internet of Services, Smart Factory and Product, Big Data and Cloud, so that the intelligent machine can analyze and use big data automatically and connect every person at every time in every where [6, 9]. The principles for a new industrial scenario to support companies identifying the possibilities are: Interoperability, Virtualization, Decentralization, Real-Time Capability, Service Orientation and Modularity, for the purpose that networks, fleets, facilities and assets can more deeply merge with the connectivity, big data and analytic of the digital world [6, 9].

### B. The Challenges to Define the Industrial Design

The acknowledged role of Industrial Design (ID for short) from industry circle is to provide design services to manufacture and ICT industry. Such services include (but not limited to) user research, product planning, concept and form innovation of products, aesthetic design, usability design and user interface design, then cooperating with engineers to complete the whole manufacture processes. Because the design related work spans from light industry to manufacture, from products to services, from tradition industry to culture industry, it is hard to define the boundary and connotation of design-related work. Though up to 2010, over 18 ID Park were set up, there is still lack of national professional level evaluation system. According to a research on a China industrial design professional evaluation system in 2011 [10], there are over 15 national occupational skill standards related to design, which are mostly constrain on a small sector of light industry and service, such as toy design, leather goods design or exhibition design; while the 29 national occupational title evaluation system are classified according to traditional disciplines branch, which is not suit to a new inter-discipline specialty like industrial design. As design is one of the driven factor transferring "China made" to "China created", it is crucial to clearly define the role of industrial designer under a whole industry chain context. From 2011, the MIIT initiated a set of researches on industrial designer occupational standard, in order that the industrial design would be professionalized and improved in proficiency level.

The definition of industrial design can't copied from another country, because it is work deeply rooted in local culture and industrialization procedure [11]. Each country has their own design features, such as the functionalism to Germany, the miniaturization to Japan, the commercial design to US, and the emotional design to Italy, which is the synthesis of their traditional and modern values, cooperation modes, social conventions, industry level and enterprises' culture [12]. In China, the emerging of industrial design is coincide with the Chinese economic reform in late 1970s, that is the symbol of a nationwide industrialization. A similar phenomenon happened in Germany in 1850s and in Japan in 1970s, but in US the popularity of ID was attributed to reduce the impact of the Great Depression and the Consumerism from Adam Smith. The experiences from industrialized countries show that the central work of industrial design is define the quality of a product or service, which are the symbol of users' lifestyle[12]. The Chinese industrial designers need to ponder carefully on our original culture and future industry vision, then plan a road map to reach a Chinese industry lifestyle. The fact is that China will never reach the quantitative modernization index like the US and EU, because of the limited resources. Thus the Chinese industry lifestyle couldn't be copied from another mode in any other country, but deeply research the sustainable needs of the Chinese [12].The bigger challenges for the Chinese industrial designers are the industrialization and post-industrialization happen at the same time and same place here, the needs from different user groups diverse like the North and South pole. What professional qualities a industrial designer expected to have in order that they could handle these challenges? It will be discussed in chapter 2.

### C. The Potential Demands of Industrial Design

Now, the product development procedure is more often viewed as an organism, such as Product Lifecycle Management and industrial ecosystem, which emphasis on the symbiosis relations of all the stakeholders [13]. From a Product Lifecycle Management perspective, the changes of industry under the context of new industrial revolution, is shown in Figure 1. In this diagram, manufacturers is replaced by makers, who emphasis personalized production and craftsmanship. With the widespread of smart small-scale production machine (3D printer as an example) and service network, it can be estimated that the Maker will boost. As it is said in the official report "Future Vision of Industry 4.0" from the German Ministry of Education and Research, in a near future every workshop, every machine will connect with each other though a industrial network, where all the information from every stage of a intelligent product is integrated together, from user's information, order requirements, shopping record, production record, processing status, shipping tracking, to user's feedback, breakdown report, maintenance record, user preferred options and left hours of the product; where the designer, maker, marketer and service would share the information all above and work together seamlessly [14]. Take the Internet of Things (IOT for short) for example. As Negroponte claimed on 2014TED, facing with a hot technology like IOT, people easily follow such ideas without enough grand vision [15]. The connection of devices to IT systems is only the first step, while the real value lies in the data that is transmitted from those devices, and the compelling business insights this data can enable[16].The fact is that the devices we need is not only "smart", but a things has their own intelligence.

Under this context, a intelligent product means the hardware who interacts with users is only a terminator who connects seamlessly together with a cloud. The designer have to plan the product and service as a whole system, as a result they will involve in all the phases of product development, and connect with users, makers, marketers and services though an intelligent manufacture platform. According to the four phases of product and service development, the future demands of industrial design could be expected as follows:

1) Strategy Level: Product = Service - A Products Service Ecosystem
2) Design Level: Personalized lifestyle Planning
3) Produce Level: Distributed, Intelligent and Eco-friendly
4) Market Level: Selling = Service = Survey
5) Service Level: Smart and Around the Clock.

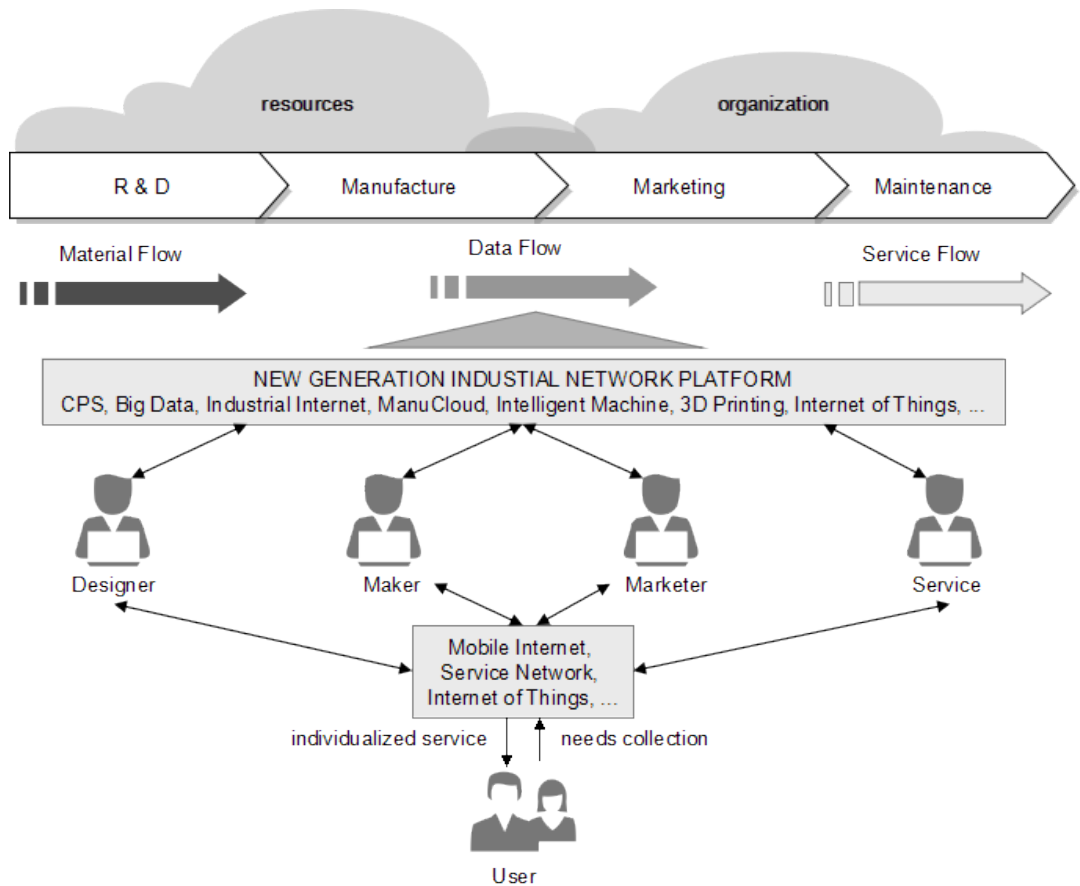

Figure 1. The diagram of products and services complex.

- Product = Service - A Products Service Ecosystem. As mentioned above, service-orientation is one feature of new industrial configuration. In consumer product sectors, the develop group of Android system from Google and iOS system from Apple both had only several hundred members, but gathered several thousand hardware manufacturers, hundred thousand engineers and million app developers together. The heavy industry can also be benefited from this principle. SANY heavy machine co. Ltd., had reduce their service cost by 60 % through a internet service platform including heavy machine remote monitor and maintenance platform, device status analysis system based on big data and self service app on a Smartphone [1]. In the future, the central task of industrial designers is to plan which needs should be satisfied, what kind of service should the company offer and what kind of bond will be built between value chain, in which the product is viewed as a hardware media of them. A long-term vision, deep understanding of user, proficiency of processing techniques, and business perspective are necessities of a industrial designer.
- R & D: Personalized lifestyle Plan. A digital product and process design system has been applied in automotive industry to cater to a fast growing diversity in product plans and individualization demands [17]. An inter-discipline data-analysis method being developed to dig useful information from a huge user Database [18]. Chinese furniture segment have already benefited from this new business model. An e-business furniture company ORDER YOUR LIFE, established in 2004, who combined O2O free online design with cloud design database, internet of things management system and digital flexible production line, had a 60 % annual growth in 2012. The resource of makers is also very powerful in R & D phases. For example, Haier Open Partnership Ecosystem had 1600 designers over the world. The prize winner Hair Dili bottles air conditioner was developed though these innovation iteration platform, where over 120 thousand user feed-backs were collected in two years, then 1 million expert and 1000 top resource reacted immediately, thus it shortened the development duration. This product created a new concept round air condition and owns over 40% market share. Under this R & D ecology, what is the different between a professional designer and a maker? The designers concentrate on analyzing huge amounts of information and organizing resources to propose a design plan for a whole product system, while the makers fulfill one or several requirements from the guideline.

- Manufacture - distributed, Intelligent and Eco-friendly. The Global Footprint Design approach used by WZL of RWTH Aachen University to support companies globally distributing their production sites and resources in order to reduce power consumption[19]. One aim of GE's industrial internet is saving 1 % fuel in aviation and power section, whose estimated value is 96 billion dollars over 15 years [6]. 3D printing, also called additive manufacturing, had widely applied in automotive enterprises, such as Festo using it for spare part management and integral part design [20]; it also applied in health-care science like dental and orthopedic surgery, as well as in light weight robotics [21]. Re-use and recycling of old parts needs also be considered in the whole production procedure [22]. Thus, the new processing techniques and environment policies are the basic knowledge of industrial designers.

- Marketing and Maintenance. The business volume of Chinese e-business giant Taobao.com in 2013 was over one trillion RMB, which was more than 10% of China GDP in the same year. Different kinds of business model, such as C2C (Customer to Customer), B2B (Business to Business), B2C (Business to Customer), O2O (Online to Offline), P2P loan (or ITFIN) had already merged into our daily life. The sales and service sectors were the start point, where ICT begin to penetrate into manufacturing. Now we have the chance to deepen and widen this trend, because single products and services are synthesizing as a facet of an intelligent service network. For example the operation information of a device will be collected and analyzed in real time, then its system will updated automatically without a periodic shutdown, and its schedule will smoothly renewed to reach higher utilization capacity. In another case, an expert system and design material database assist the user to design their own products, then the information of the order automatically distribute to purchase, stock, produce and logistics, and the processing information will transparently shared with user so that they can change the design spec at any processing phases. In this scenario, every user has intimate connection with their own product and experiences personalized service in a whole design and production procedure [14].

## II. REDEFINE THE ROLE OF INDUSTRIAL DESIGNER

Since the 9th national five-year plan, China began to transit its economic development mode to an environmental friendly, resources efficiently, and sustainable one. The demand of transiting "China-made" to "China-create" boosted the need of ID professions. For a long time, the boundary and definition of industrial design are unclear. In 2010, Guangdong province initiated a pilot project tried to propose an ID designer occupation standard, but this project focused mostly on product design in light segment of manufacturing industry [ID project]. Though in ICT the demand of new emerging design professions, such as user research, user experience, interaction design, user interface design, usability test increases since 2000 [ID], their job requirements are mostly depend on companies conventions. In 2011, The China MIIT entrusted the department of Industrial Design, Xi'an Jiaotong University (ID XJTU for short) to research the occupational standard of industrial design. In this research, the professional qualities expected in the two ID professions, product designer and human-computer interface designer were both investigated, so as to define the role of ID designer comprehensively.

For the sake of conceiving a comprehensive framework of ID designer professional level, the job requirements were reviewed or interviewed from management groups in foreign and domestic companies ( see Table 1). Then 135 managers and 114 leaders of ID department in local companies and universities were surveyed by a questionnaire study to verify the validity of the framework.

TABLE I.   THE STATISTICS OF INVESTIGATED COMPANIES AND UNIVERSITIES

| Type of Organizations | Product Designer Demands | Human-Computer Interface Designer Demands |
|---|---|---|
| foreign companies reviewed | 20 | 18 |
| domestic companies interviewed | 41 | 15 |
| domestic companies surveyed | 135 | / |
| domestic universities surveyed | 114 | / |

The professional level framework covers three main aspects, that is humanity qualities, job responsibilities and professional abilities. Because some of the job responsibilities of product design and HCI design are in common, this aspect could be subdivided into 3 parts: Job Responsibilities in common, Product Requirements and HCI Requirements. Thus, the framework consists of five main factors, as shown in Figure 2. The 35 secondary factors covered all the qualities required for junior, mid-level and advanced industrial designer.

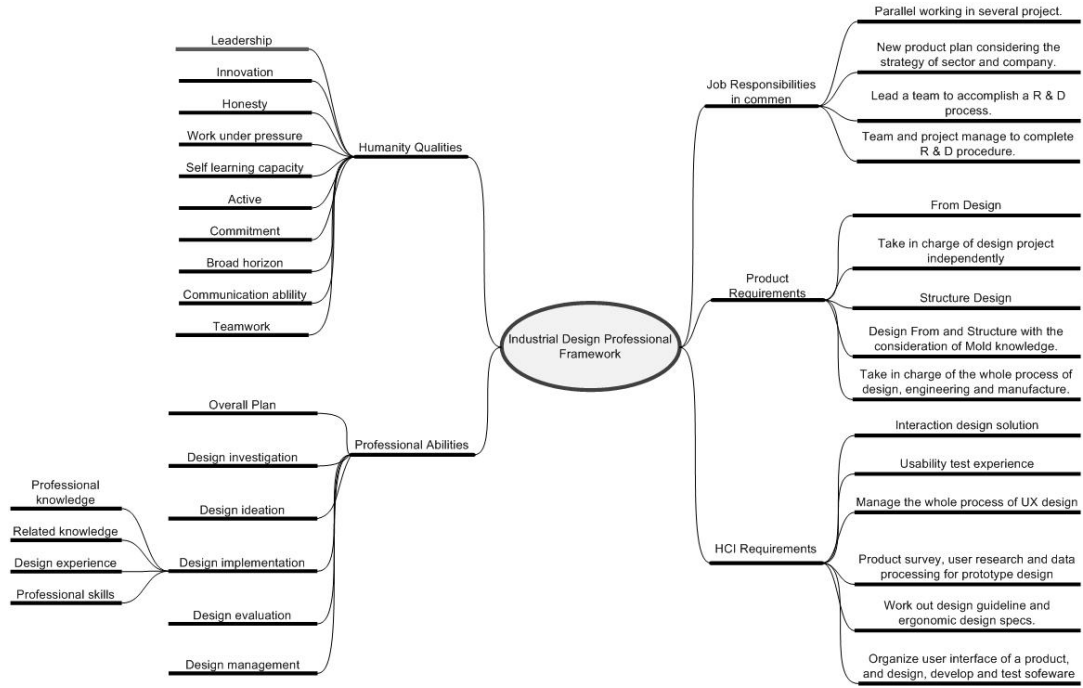

Figure 2.  The professional level framework of Industrial Design.

## A. Humanity Qualities

The humanity qualities refer that the characters reflecting culture on a single person, that is concentrated on the core value and instrumental values. The core value of a industrial designer should be love and kindness, for the commitment of industrial designer is planning a future culture and lifestyle [11]. The instrumental values, which means the methods to achieve core value, related with ten factors like Teamwork, Broad horizon, Communication abilities, Commitment. The first eight humanity qualities are the baseline of a designer, and the ninth is for a mid-level designer. While an advanced designer should have all the qualities.

TABLE II.  THE HUMANITY QUALITIES FACTORS EXPLANATION

| Secondary factors | Explanation |
|---|---|
| Teamwork | a. Team-centered, team-integrated, not self-centered;<br>b. Good coordinator, good assistant, and problem solver in production.;<br>c. Have good internal and external cooperative relation , and pursuing common development with internal and external partners based on a win-win goal. |
| Communication abilities | a. Good listener;<br>b. Competent speaker;<br>c. Getting resources and support from outside through communication. |
| Broad horizon<br>Broad horizon | a. Suggest design goal, user requirements, market strategies, processing methods, quality and eco-standard, from a whole company and segment perspective;<br>b. Consider and explore the requirements of future design, with the consideration of future way of survival, and strategies of company. |
| Commitment | a. Plan future lifestyle with love and kindness;<br>b. Have sense of social responsibility and occupation moral, and responsible for the occupation activity in long-term;<br>c. Dutiful. |
| Active | a. Think independently, and take work actively. |
| Self learning capability | a. Lifelong learning; willing to try new things, willing to share knowledge;<br>b. Good at learning new techniques, good at searching new methods, willing to improve operational system in work. |
| Work under pressure | a. High EQ; resolve the causes of work pressure;<br>b. keep calm in stress, and ease the confrontation and conflict. |
| Honesty | a. Keep promise, match word to deed, adhere to original design, and respect IP rights of others. |
| Innovation | a. Explore actively to solve the unknown problems, with great enthusiasm;<br>b. Able to plan and develop new products. |
| Leadership | a. Able to guide and manage design teams, and set up design principles. |

## B. Professional Abilities

The professional abilities refer to the professional abilities and procedural knowledge required in each product development phases. A single design project could be separated into 4 phases: investigation, ideation, implementation and evaluation. Before the very beginning of a project, an overall plan should initiated to identify an overall strategy. The design management penetrates all stages of a R & D process.

TABLE III. THE PROFESSIONAL ABILITIES FACTORS

| Secondary factors | | Explanation |
|---|---|---|
| Overall plan | | a. Able to planning a project considering user needs, company strategy, design strategy, produce strategy, market strategy. |
| Design investigation | | a. Able to conduct segment survey, product strategy survey, feasibility analysis, user investigation and market survey; b. Master systematical methods of investigation and data analysis, ensuring the validity and reliability of surveys; c. Conclude user model and design guideline. |
| Design ideation | | a. Identify the product strategy, design goal and solutions based on survey results, user model and design guidelines; b. Product concept manifest sustainable values, morals, lifestyle and mode of production. |
| Design implementation | a. Professional knowledge | The foundations of product development, including: design theory like sociology, psychology, aesthetics, ergonomics, semiotics; and engineering knowledge like material, processing, structure, mold. |
| | b. Related knowledge | A wide range of knowledge, such as segment related techniques, policies, rulers, marketing, management and QC. |
| | c. Design Experience | Design project had put into market, and organize and lead a team to operate the project independently. |
| | d. Professional skill | Proficient at CAD and design expression, including sketching, engineer drawing, rendering, modeling and prototyping. |
| Design evaluation | | a. Test and verify designs; b. Accumulate design experience to compile design guidelines; c. Adjust product strategy to improve product. |
| Design management | | a. Constitute design strategy, design methodology and design system, and design team building and personnel training; b. Able to operate a project, such as project planning, team organizing, coordinating and problem solving, time planning, cost control, testing and QC. |

## C. Job Responsibilities in Common

The job responsibilities are the work content undertaken by different level of designer. No matter what kind of designer they are, the following contents are involved in daily work of an advanced design.

TABLE IV. THE JOB RESPONSIBILITIES IN COMMON

| Secondary factors | Explanation |
|---|---|
| Lead a team to accomplish a R & D process | a. Have practical experience on design project; b. able to push the team go forward. |
| Parallel working in several project | a. Able to coordinate different design project; b. well-organized and fast reaction; c. solve problem quickly. |
| New product plan considering the strategy of the segment and company | a. Have high vision to recognize the real needs of users; b. predict the future trends of industry. |
| Team and project manage to complete R & D procedure | a. Able to organize and establish design team and manage project to achieve company vision. |

## D. Product Requirements

This group of work contents are closely related to develop a product - a physical hardware.

TABLE V. THE PRODUCT REQUIRMENTS

| Secondary factors | Explanation |
|---|---|
| From design | Able to design an extrinsic feature of a product according to different user aesthetics. |
| Take in charge of a design project independently | Able to responsible to a product design independently. |
| Structure design | Able to design assembly and parts with the accordance of product function. |
| Design form and structure with the consideration of mold knowledge | Have rich experience on processing technology to ensure the quality of a product. |
| Take in charge of the whole process of design, engineering and manufacture | Able to organize a whole product development process, and cooperate with another work role to solve problem during production. |

## E. HCI Requirements

This group of work contents aim to define human-computer interfaces, as well as human computer interaction flows.

TABLE VI. THE HCI REQUIRMENTS

| Secondary factors | Explanation |
|---|---|
| Interaction design solution | a. Able to convert an interaction concept to a design solution. |
| Usability test experience | a. Have usability testing experience; b. Able to conduct a usability test. |
| Product survey, user research and data processing for prototype design | a. Able to design a user interface demo according to the result of design investigation and user model. |

| Secondary factors | Explanation | |
|---|---|---|
| Manage the whole process of UX design | a. | Able to initiate a UX project; |
| | b. | Familiar with UX principles; |
| | c. | Able to evaluate UX level of a product. |
| Work out design guideline and ergonomic design specs | a. | Have knowledge on psychology and sociology; |
| | b. | Have knowledge on usability; |
| | c. | Have knowledge on ergonomics; |
| | d. | Have knowledge on computer system. |
| Organize user interface of a product, and design, develop and test software | a. | Able to suggest a "user-centered" HCI ideation; |
| | b. | Able to design a logic of a software; |
| | c. | Able to develop a software; |
| | d. | Able to testing a software. |

According to the Chinese industry vision "China Manufacture 2025", the intelligent devices are the key to achieve the manufacturing innovation. This trend indicates that the division of product designers and HCI designers will gradually vanish, and the task of designers is to rebuild product concept for an innovative, intelligent, quality and green lifestyle. It is including: way of work, way of production, way of life, way of recreation, way of transportation, way of entertainment, way of connection, and way of communication.

## III. THE EDUCATION MODEL OF INDUSTRIAL DESIGNER

The new industrial revolution that we are facing means that the way of survival is changing. Planning products and services to achieve a sustainable coexisting lifestyle and culture is the central task and obligation of an industrial designer [11]. In order to help students preparing for the challenges from future, the demanded qualities of future designers need to be integrated into the ID education model. The purpose of high education should be "human centered" and achieving a "Complete Development of a Person" [23], that is all the qualities and abilities for his/her future development should be inspired thoroughly though a education plan. To ensure this purpose, the following questions should be answered:

1. "How do we know the graduate able to take full charge of their job?"

The education model of ID students should be in accordance with the professional level framework of ID designers.

2. "Will they have sustainable careers?"

The next three decades will be a dramatic changing era for China industry, and there is no experience or model to follow. So the knowledge we teach today will not solve the future problems of the graduates. Li leshan found that cultivating the humanity qualities and abilities would granted students the competencies for exploring new knowledge and solutions [24].

3. "What is the 'Complete Development of a Human Being'?"

1) A fully developed person has five hierarchies:
2) A individual with healthy physical and mental status;
3) A family member living in harmony ;
4) A profession fully in charge of work;
5) A civilian willingly to take social duties;
6) A nation passing on his own culture. [25]

These five aspects of a student can attribute to three main qualities, namely Personality, Humanity and Ability. The International Council of Societies of Industrial Design evaluates a world-class university according to following factors:

1) Leadership and Initiative;
2) Teamwork ability;
3) Innovative thinking ability;
4) Communication ability;
5) Fast adaptation ability;
6) Able to recognize the factors causing social changes;
7) Aesthetic ability;
8) Morality.

All the factors above evaluate the humanity qualities but not the knowledge levels. Li leshan added an **international perspective** factor and established a Personality, Humanity and Ability education model (PHA for short) for industrial design [25]. The PHA education model was authorized by ICSID in 2003 and commented as "corresponds to most modern program of ID education in the world" by the Executive Board Member of ICSID.

### A. PHA Education Model

As shown in Figure 3, in PHA education model, personality is the dominant factor, humanity and ability are the supporting factor. This model includes all the factors required by ICSID.

The personality means that the students respect self and others, distinguish between right and wrong, have determine-minded and good mental health. The humanity covers 3 secondary factors, values, that are kindness, industrious, innovative, reasonable, efficient and quality; morality, that are self commitment, family commitment, occupation commitment and social commitment; way of act, that the students should act extroverted and community-oriented.

How can we improve the humanity qualities of the students? One method is "teaching by discussing", which indicate that students need to find and solve problems themselves by discussing, instead of asking an answer from teachers. Why students could improve their humanity qualities though discussion? Firstly, the prerequisite of effective discussion are the kindness of team members, restraining their egocentricity, acting together, forgiveness and justice, as well as the will to learn and improve together. Secondly, in a discussing, every student is required suggesting rather than complaining. Everyone joining in a discussion is bound to share their ideas and communicate with others, thus to be open-mind and friendly. Thirdly, it gives them the chance to learn different cognitive methods from each other and cultivate their own way of recognizing [26]. Moreover, during humanity courses, students required to conduct several research projects, like "college student value investigation", "family history and culture research", "color preference investigation", "high quality product investigation". These project give them the good understanding of modern culture in China."China-made" is complained as lack of innovation, because most of the researchers and designers lack of in depth pondering on the human nature and needs, thus lack of the deep level understanding of quality of life.

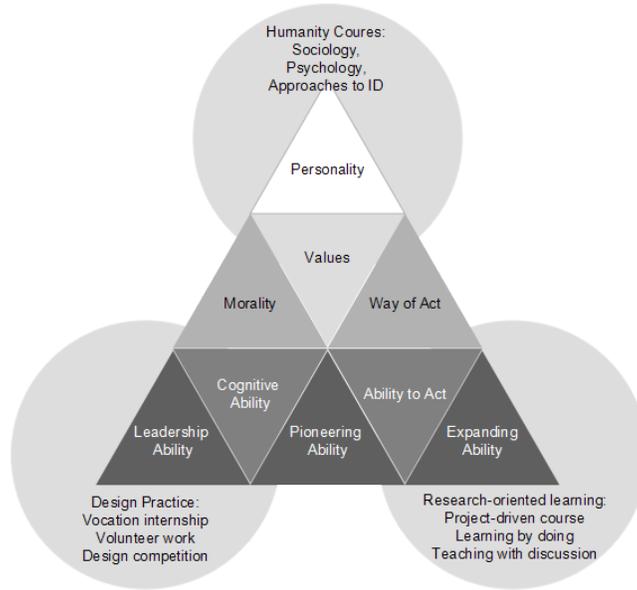

Figure 3. The professional level framework of Industrial Design

The ability contains two aspects, that is psychological ability and innovative ability. The psychological ability includes ability to act, meaning the able to act independently, such as set a goal, make a plan, perform an action and evaluate the result; and cognitive ability, meaning the ability to cognize, such as observing, heuristic thinking, memorizing, understanding, communicating, problem-finding and problem-solving. Innovative ability means leadership ability, pioneering ability and expanding ability, such as able to propose a new conception, categorize information, definition ability, make connections between different concepts. "Leaning by doing" is the only way to promote the abilities, because the abilities are procedural knowledge which can only learned though practice other than understanding and memorizing [27]. The practice means autonomic learning, which contain initiate a project independently, find resources independently, design a research about this project independently, plan this project independently, make a design originally, test the design independently and evaluate the project independently. In scientific research and product development, the most important ability is identifying and determining a project, a training of which is missing in Chinese high education program [28]. That is why the Chinese products are short of original ideation and design. Furthermore, students required to practice in companies every vocation, in order to enrich their experience on processing technologies and professional ways of working, and more important is to broaden their horizon on the society and industry. Each graduate from ID department in Xi'an Jiaotong University works in about 7 different companies in total over 48 weeks, which equal to one year work experience.

Continual improvement and modification of the PHA model correspond to the fast-changing industry condition and new requirement from companies. Deep social practice, keeping contact and cooperating with alumni and enterprises are the key to improve.

*B. A Survey about the Education Quality*

The occupation type, annual income and self-initiated business are well accepted indexes referring the education quality. Thus, from April 29 to May 12, 2015, a real name survey was conducted on 48 Alumni from ID XJTU about their income levels, their first occupation, current occupation, self-initiated business, as well as their suggestions about the curriculum and the PHA education model.

78% of the participants replied that the PHA education model gave them good help in personal development, such as good self-learning ability, willing to cooperate actively, a sound personality, commitment for good work, refraining egocentricity. Someone mentioned that the internship experience help them easily find a good job, while the humanity quality help them easily handle every challenge in work and get more promotion chance. Another said that PHA education is crucial to initiating his own business, when he solved a complex problem and coordinate with his partners.

The graduation time spans from 2003 to 2014, including every year of graduate (except 2006), as shown in Figure 4. 58.3% of the participants were graduated in last five years.

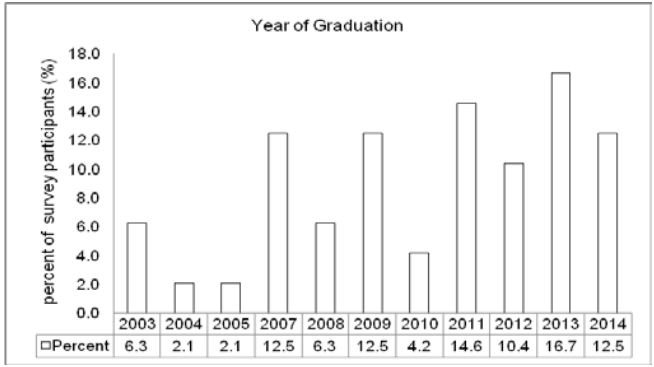

Figure 4. Percentage of the participants in different graduate year. The first groups of bachelors and masters graduated in 2003.

According to a survey report from Mycos research [29], in 2015 the average salary of bachelors is 3694 yuan/month, of masters is 5590 yuan/month. The graduates have the highest salary in Beijing, Shanghai, Guangzhou and Shenzhen, and the average bachelor salary there is 4364 yuan/month, while the average master salary there is 6503 yuan/month. 46.9 % of the graduates from ID department in Xi'an Jiaotong University had over 8000 yuan/month, 34.1% of the graduates had about 10,000 yuan/month. It shows that the graduates from ID XJTU have more higher - level occupations.

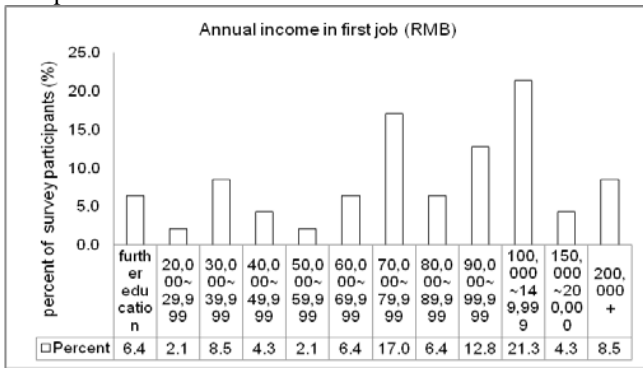

Figure 5. Annual income of the graduates' first job distribution. 34.1% of the graduates had over 100,000 annual salary.

76.6% of the graduates from ID XJTU had income now about 10,000 yuan/month, while 59.6% of the graduates had about 20,000 yuan/month. Another nationwide survey from Mycos research said the bachelors (who graduated in 2010) after 3 years earned 5962 yuan/month in average, while the bachelors (who graduated in 2009) after 3 years earned 5350 yuan/month in average [30]. It indicates that the ID XJTU graduates got promotion faster.

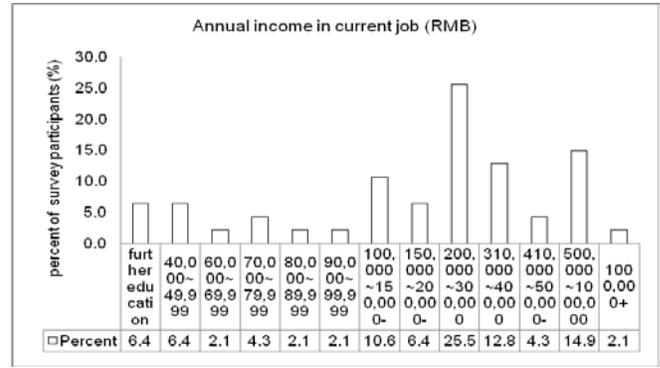

Figure 6. Annual income of the graduates' current job distribution.

The types of first occupation indicates that 91% of the graduates form ID XJTU work in ICT and manufacturing segment, and 71% worked as designers, developers or managers, as shown in Figure 7. In 5 years over 40% of the graduates have promoted to management layer, as shown in Figure 8. The result proves the statement of the graduates that PHA give them more chance to glint in their work and get promotion faster.

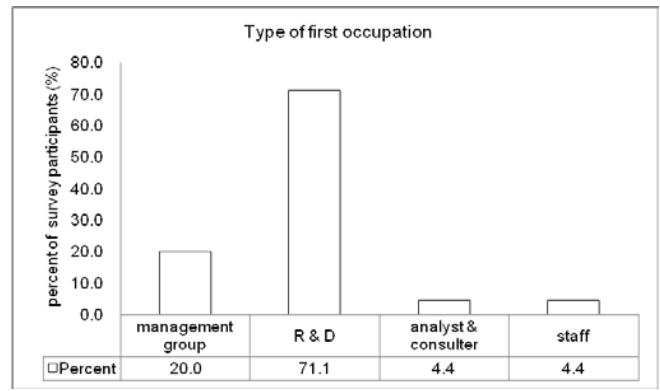

Figure 7. The type of first occupation of the graduate shows that 90% of them work as designer or management.

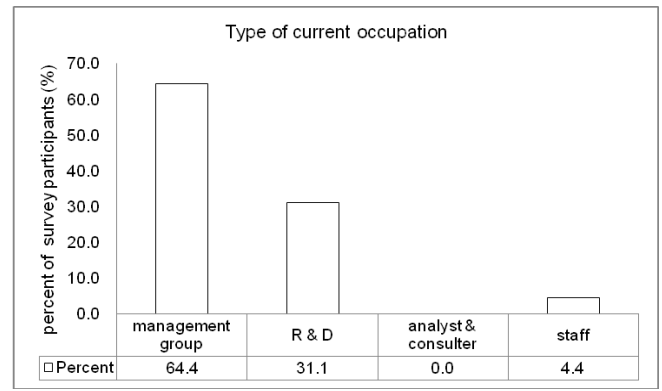

Figure 8. The type of current occupation of the graduate shows that more than 40% have already in management group or open their own business.

The human resource website Zhaopin.com released the result of employment survey on 2015 graduates [31], which shown 6.3% of them will choose initiate their own business.

29.5% of the graduates from ID XJTU had opened their own business (see Figure 9). This result is also corresponds to the conclusion of type of current occupation.

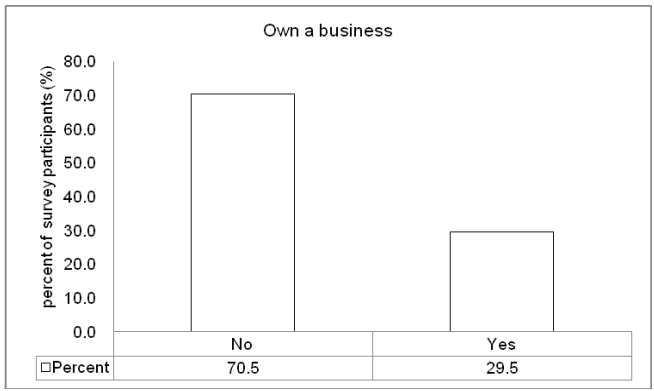

Figure 9. The percentage of graduate initiate their own business.

## IV. CONCLUSIONS

1. **The new industrial reform gives greater developing chance to industrial design than ever, as well as tougher challenges.** The manufacturing and ICT blueprint "China Manufacturing 2025" suggests a vision of manufacturing segment, which will be innovative, high-quality, intelligent and green, where the innovation factor are the foremost. The "Internet +" is the trial started in ICT industry to penetrating into the future. This dynamic time requires clear vision and broad horizon of industrial designers to explain the factors causing the social change and to predict the future main stream, as well as the sensitivity to new technology, new material and segment trend.

2. **The foremost ability of an industrial designer is the humanity qualities, which will equip them for the fast changes, new challenges and complex tasks.** A correct value judgment help them avoiding pursuing short-term profits and resisting temptations, so that they can insist on a sound decision. The basic professional ability for the future designer is he/she can do well on the project which they haven't learned, haven't seen, haven't work with it. The inter-disciple and inter-segment ability are also the necessities.

3. **These required abilities can be inspired though a PHA education model.** In this model, the personality and humanity are emphasized and judged in every professional course. The purpose of education is not convey knowledge but achieve fully development of each student. "Teaching by discussing" and "learning by doing" are the chief teaching methods. A continuous improvement of education model is needed to meet the new ends – new challenges and chances.

ACKNOWLEDGMENT

One of us ( Meng LI ) thanks Sebastian Degen from German Department, Xi'an Jiaotong University, because of his kind help on German references explanation and advises. This paper is funded by Shaanxi innovation expert education experiment area project.